\documentstyle[aps,prb,epsf, multicol]{revtex}
\begin{document}
\draft
\newcommand{\bn}{{\bf n}}
\newcommand{\bp}{{\bf p}}
\newcommand{\br}{{\bf r}}
\newcommand{\bq}{{\bf q}}
\newcommand{\bj}{{\bf j}}
\newcommand{\bE}{{\bf E}}
\newcommand{\eps}{\varepsilon}
\newcommand{\la}{\langle}
\newcommand{\ra}{\rangle}
\newcommand{\cK}{{\cal K}}
\newcommand{\cD}{{\cal D}}
\newcommand{\mybeginwide}{
    \end{multicols}\widetext
    \vspace*{-0.2truein}\noindent
    \hrulefill\hspace*{3.6truein}
}
\newcommand{\myendwide}{
    \hspace*{3.6truein}\noindent\hrulefill
    \begin{multicols}{2}\narrowtext\noindent
}

\title{
  Nonlinear conductivity of diffusive normal-metal contacts\\
  \rm   Physics Letters A {\bf 189}, 134 (1994)$^*$
}

\author{
               K.\ E.\ Nagaev
}
\address{
  Institute of Radioengineering and Electronics,
  Russian Academy of Sciences, Mokhovaya ulica 11, 103907 Moscow,
  Russia}
\date\today
\maketitle
\bigskip
\begin{abstract}
Metal microbridges with a high impurity content and shorter than the
energy relaxation length are considered. Their conductance is calculated
with allowance made for the Coulomb electron-electron interaction. It is
shown that nonequilibrium electrons in the microbridges gives rise to a
nonlinear current-voltage characteristic.
\end{abstract}
\begin{multicols}{2}
\narrowtext

\nopagebreak
It is common knowledge that tunnel junctions between disordered metals
show nonlinear conductance at low temperatures (the so-called zero-bias
anomaly), which results from the electron-electron interaction.\cite{1}
On the other hand, short microbridges possess meny properties of tunnel
junctions. Therefore, it is interesting to investigate their nonlinear
properties at low temperatures.

Nonlinear conductivity of weakly disordered metals at low temperatures was
investigated both ex\-pe\-ri\-men\-tal\-ly\cite{2,3} and
the\-ore\-ti\-cal\-ly\cite{4} in a
number of papers. In these papers, the nonlinear behavior resulted from
the heating of the electron gas by the current and from the
temperature-dependent correction to the conductivity arising either from
weak localization or electron-electron interaction effects. The resulting
current-voltage characteristics were influenced, however, by several
parameters describing the heat transfer from the electron gas in a massive
sample, e.g., by the electron-phonon coupling constant and the acoustic
transparency of the sample boundary. Therefore it was difficult to compare
the theoretical results with theoretical predictions. For metal
microbridges, the situation is different because the heat transfer is
determined by the diffusion of hot electrons into the banks and no
additional parameters are required.

In this paper we investigate the influence of Coulomb exchange
electron-electron interactions on the current-voltage characteristics of
disordered-metal microbridges shorter than the energy relaxation length.
The characteristic energy scale in this case is the largest of the
quantities $eV$ and $kT$, wher $V$ is the bias voltage and $T$ is the
temperature. Therefore the electron distribution function is essentially
nonequilibrium at low temperatures $kT \ll eV$, and one may expect a
nonlinear current-voltage dependence.

Consider a metal microbridge connecting two massive banks with the $x$
axis directed along the contact. The microbridge length $L$ is assumed to
be larger than the electron elastic mean free path and the microbridge
width. Yet, $L$ is assumed to be shorter than the energy relaxation
length.

In studying the current-voltage characteristics of the contact under
nonequilibrium conditions, the Keldysh diagrammatic technique\cite{5} may
be conveniently used. In this technique, three types of Green functions
are used,
$$
 G^R(1,2)
 =
 i\theta(t_2 - t_1)
 \la
   \psi(1)\psi^{+}(2)
   +
   \psi^{+}(2)\psi(1)
 \ra,
$$ $$
 G^A(1,2)
 =
 -i\theta(t_1 - t_2)
 \la
   \psi(1)\psi^{+}(2)
   +
   \psi^{+}(2)\psi(1)
 \ra,
$$ $$
 G^K(1,2)
 =
 -i
 \la
   \psi(1)\psi^{+}(2)
   -
   \psi^{+}(2)\psi(1)
 \ra.
$$
In the absence of the electron-electron interaction, the impurity-averaged
retarded and advanced Green functions are given by the well-known
expressions
\begin{equation}
 G^{R(A)}(\eps,\br, br')
 =
 \int\frac{d^3p}{ (2\pi)^3 }
 \frac{
   \exp[
      i\bp(\br - \br')
   ]
 }{
   \eps - p^2/2m \pm i/2\tau
 },
\label{1}
\end{equation}
where $\tau$ is the elastic relaxation time. The Green function $G^K$
obeys the integral equation
$$
 G^K(\eps, \br, \br')
 =
 \frac{1}{2\pi N\tau}
 \int d^3 r_1\,
 G^R(\eps, \br, \br_1)
$$ \begin{equation}
 \times
 G^K(\eps, \br_1, \br_1)
 G^A(\eps, \br_1, \br'),
\label{2}
\end{equation}
where $N$ is the electron density of states at the Fermi level. By setting
$\br = \br'$ in (\ref{2}), one may obtain a diffusion equation for the
Green function $G^K(\eps, \br, \br')$ (see Ref. \cite{6})
\begin{equation}
 D\nabla^2
 G^K(\eps, \br, \br)
 =
 0,
 \qquad
 D
 =
 \frac{1}{3}
 v_F^2\tau.
\label{3}
\end{equation}
In the case of massive banks, the electron distribution at the ends of the
microbridge is equilibrium and therefore the boundary conditions for
Eq.~(\ref{3}) are
\begin{eqnarray}
 \left.
  G^K(\eps, \br, \br)
 \right|%
 _{x = -L/2}
 =
 -2\pi N
 [
  1 - 2n(\eps - eV/2)
 ],
\nonumber\\
 \left.
  G^K(\eps, \br, \br)
 \right|%
 _{x = L/2}
 =
 -2\pi N
 [
  1 - 2n(\eps + eV/2)
 ],
\label{4}
\end{eqnarray}
where $n(\eps)$ is the Fermi distribution function and $V$ is the voltage
drop across the contact. The current flowing through the contact may be
expressed in terms of the Green function $G^K$ via the formula
\begin{equation}
 j_x
 =
 -\frac{e}{2m}
 \int d\eps
 \left(
   \frac{\partial}{\partial x}
   -
   \frac{\partial}{\partial x'}
 \right)
 \left.
  G^K(\eps, \br, \br')
 \right|%
 _{\br = \br'}.
\label{5}
\end{equation}
\begin{figure}
\centerline{
  \epsfxsize7.5cm
  \epsffile{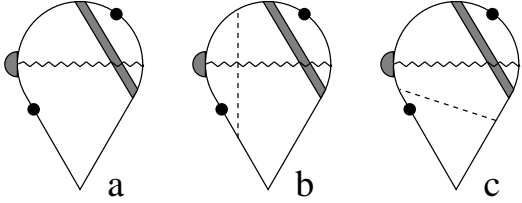}
}
\caption{
 The diagrams describing the current to first order of the
 electron-electron interaction.
}
\label{FIG1}
\end{figure}
\noindent
The corrections to the current from the electron-electron interaction are
represented by the three diagrams shown in Fig. 1 and their complex
conjugates. In these diagrams, solid lines denote the advanced and
retarded electron Green functions, the black circles denote the Green
functions $G^K(\br, \br)$. The single dashed lines represent the
impurity-potential correlator $(2\pi N\tau)^{-1}\delta(\br - \br')$. The
shaded rectangles denote the impurity-averaged two-particle Green functions
\begin{equation}
 P(\omega, \br, \br')
 =
 \la
   G^A(\eps + \omega, \br, \br')
   G^R(\eps, \br', \br)
 \ra_{imp},
\label{6}
\end{equation}
which obey the equation
$$
 (
  i\omega + D\nabla^2
 )
 P(\omega, \br, \br')
 =
 -2\pi N\delta(\br - \br'),
$$ \begin{equation}
 \left.
  P(\omega, \br, \br')
 \right|%
 _{x = \pm L/2}
 =
 0.
\label{7}
\end{equation}
The shaded semicircles denote the impurity-renormalized electron vertex
$(2\pi N\tau)^{-1}P(\omega, \br, \br')$. The electron-electron interaction
is represented by the wavy lines. Taking into account the Debye screening,
one obtains the following equation for the retarded potential of the
electron-electron interaction,
$$
 D\nabla^2
 V^R(\omega, \br, \br')
 =
 N^{-1}
 (
  -i\omega + D\nabla^2
 )
 \delta(\br - \br'),
$$ \begin{equation}
 \left.
  V^R(\omega, \br, \br')
 \right|%
 _{x = \pm L/2}
 =
 0.
\label{8}
\end{equation}
The advanced potential $V^A$ is the complex conjugate of $V^R$.

These diagrams result in a contribution to the current density in the form
$$
 \bj_1(\br)
 =
 2(2\pi)^{-5} eDN^{-3}
$$ \vspace{-5mm} $$
 \times {\rm Im}\,
 \int d\eps  \int d\omega
 \int d^3r_1 \int d^3r_2\,
 G^K(\eps, \br, \br)
$$ \vspace{-5mm} $$
 \times P(-\omega, \br, \br_1)
 G^K(\eps - \omega, \br_1, \br_1)
 V^R(\omega, \br_1, \br_2)
$$ \vspace{-5mm} \begin{equation}
 \times\nabla_{\br}
 P(-\omega, \br_2, \br).
\label{9}
\end{equation}
To take into account the current-conservation law, this current density
should be averaged over the contact length.

Substitute the solutions of (\ref{3}), (\ref{7}), and (\ref{8}) into
(\ref{9}). In the low-temperature limit $kT \ll eV$, the Fermi
\begin{figure}
 \epsfxsize8cm
 \epsffile{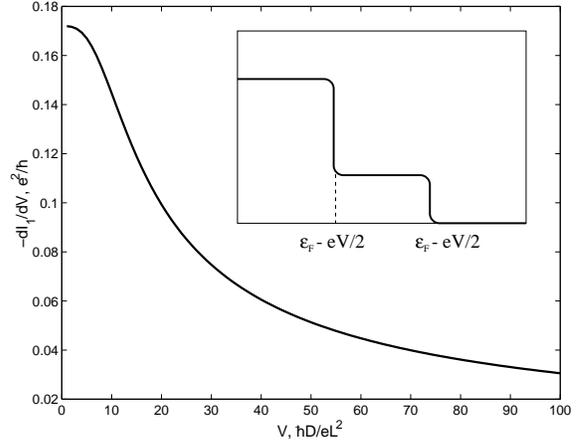}
 \caption{
   The differential conductance versus voltage. The inset shows the
   electron distribution function at an arbitrary point inside the
   contact.
 }
\label{FIG2}
\end{figure}
\noindent
distribution function is step-like and the integration over $\eps$ may be
easily performed. The integration with respect to $\omega$ in (\ref{9})
may be conveniently replaced by the integration with respect to a
dimensionless quantity $y$, which is related to $\omega$ by $y^2 =
(L^2/2D)\omega$. Therefore, the correction to the current flowing through
the contact may be presented in the form
\begin{equation}
 I_1
 =
 -\frac{16}{\pi}
 \frac{eD}{L^2}
 \left(
   s^2
   \int\limits_s^{\infty} dy
   \frac{ Q(y) }{ y^3 }
   +
   \int\limits_0^s dy
   \frac{ Q(y) }{ y }
 \right),
\label{10}
\end{equation}
where $s = L(eV/2\hbar D)^{1/2}$ and
$$
 Q(y)
 =
 \{
   (1/4)y
   [
     \sinh(2y) - \sin(2y)
   ]
$$ \vspace{-7mm} $$
   -
   y^{-1}
   [
     \sinh(2y) + \sin(2y)
   ]
$$ \vspace{-7mm} $$
   +
   2y^{-1}
   [
     \sinh(y)\cos(y) + \sin(y)\cosh(y)
   ]
 \}
$$ \vspace{-7mm} \begin{equation}
 \times
 [
   \cosh(2y) - \cos(2y)
 ]^{-1}.
\label{11}
\end{equation}
The corresponding differential resistivity versus voltage curve is shown
in Fig.~2. At high voltages $eV \gg \hbar D/L^2$, $I_1 \approx -(8/\pi)
(DeV/2\hbar L^2)^{1/2}$. Therefore, the voltage dependence of resistivity
in this voltage range has the shape characteristic of the temperature
dependence of resistivity in one-dimensional conductors.\cite{1} This is
quite natural, because the width of the drop in the electron distribution
function is now determined by the applied voltage and not by the
temperature. Therefore, the characteristic momentum transfer in (\ref{9})
is of the order of $(eV/\hbar D)^{1/2}$. In the opposite case of small
voltages $eV \ll \hbar D/L^2$, the momentum transfer is limited by
$\hbar/L$. Therefore, the conductivity tends to a constant value, i.e.,
$I_1 = -0.17(e^2/\hbar)V$. Hence, increasing voltage gives rise to a
cross-over from a zero-dimensional case to a quasi one-dimensional case.

Note that the effect of the electric field is not reduced to a simple
heating of the electron gas to some effective temperature. Instead, the
electron distribution function consists of two subsequent steps positioned
at $\eps_F \pm eV/2$ (see the inset in Fig. 2) and therefore has an
essentially non-Fermian shape. In principle, the nonequilibrium electron
distribution may also affect the phase-breaking time and therefore the
weak localization contribution to the conductance.\cite{4} However, weak
localization corrections may be suppressed by a sufficiently strong
magnetic field, and the universal conduction fluctuations\cite{6,7} may be
eleminated by averaging the results over a number of samples.

The author is grateful to M.E. Gershenson, G.A. Ovsyannikov, and
A.V. Zaitsev for discussing the results. This work was partially supported
by the Soros International Science Foundation and the Russian Foundation
for Fundamental Research.

\end{multicols}
\end{document}